\documentclass[sn-mathphys-num]{sn-jnl}

\usepackage{graphicx}%
\usepackage{multirow}%
\usepackage{amsmath,amssymb,amsfonts}%
\usepackage{amsthm}%
\usepackage{mathrsfs}%
\usepackage[title]{appendix}%
\usepackage{xcolor}%
\usepackage{textcomp}%
\usepackage{manyfoot}%
\usepackage{booktabs}%
\usepackage{algorithm}%
\usepackage{algorithmicx}%
\usepackage{algpseudocode}%
\usepackage{listings}%

\theoremstyle{thmstyleone}%
\theoremstyle{thmstyletwo}%

\theoremstyle{thmstylethree}%

\begin{document}

\title[Article Title]{Conditional Diffusion-based Parameter Generation for Quantum Approximate Optimization Algorithm}


\author[1]{\fnm{Fanxu} \sur{Meng}}
\equalcont{These authors contributed equally to this work.}

\author*[1]{\fnm{Xiangzhen} \sur{Zhou}}\email{xiangzhenzhou@njtech.edu.cn}
\equalcont{These authors contributed equally to this work.}

\author[2]{\fnm{Pengcheng} \sur{Zhu}}

\author[3]{\fnm{Yu} \sur{Luo}}


\affil*[1]{\orgdiv{College of Artificial Intelligence}, \orgname{Nanjing Tech University}, \orgaddress{\street{Puzhu South Road}, \city{Nanjing}, \postcode{211800}, \country{China}}}

\affil[2]{\orgdiv{College of Information Engineering}, \orgname{Taizhou University}, \orgaddress{\street{Jichuan East Road}, \city{Taizhou}, \postcode{225300}, \country{China}}}

\affil[3]{\orgdiv{College of Computer Science}, \orgname{Shaanxi Normal University}, \orgaddress{\street{Chang An Road}, \city{Xi'an}, \postcode{710062}, \country{China}}}




\abstract{The Quantum Approximate Optimization Algorithm (QAOA) is a hybrid quantum-classical algorithm that shows promise in efficiently solving the Max-Cut problem, a representative example of combinatorial optimization. 
However, its effectiveness heavily depends on the parameter optimization pipeline, where the parameter initialization strategy is nontrivial due to the non-convex and complex optimization landscapes characterized by issues with low-quality local minima. 
Recent inspiration comes from the diffusion of classical neural network parameters, which has demonstrated that neural network training can benefit from generating good initial parameters through diffusion models. However, whether the diffusion model can enhance the parameter optimization and performance of QAOA by generating well-performing initial parameters is still an open topic. Therefore, in this work, we formulate the problem of
finding good initial parameters as a generative task and propose the initial parameter generation scheme through dataset-conditioned pre-trained parameter sampling. Concretely, the generative machine learning model, specifically the denoising diffusion probabilistic model (DDPM), is trained to learn the distribution of pre-trained parameters conditioned on the graph dataset. Intuitively, our proposed framework aims at effectively distilling knowledge from pre-trained parameters to generate well-performing initial parameters for QAOA.
Compared to random parameter initialization, experiments on various-sized Max-Cut problem instances consistently show that our conditional DDPM is capable of improving the approximation ratio by as much as $14.4\%$, $11.0\%$, $11.4\%$ and $7.49\%$, $8.31\%$, $6.08\%$ on average for random, regular, and Watts-Strogatz graphs, respectively. Additionally, the experimental results also indicate that the conditional DDPM trained on small problem instances can be extrapolated to larger ones, improving the approximation ratio by up to $28.4\%$ and $12.1\%$ on average, thus demonstrating the extrapolation capacity of our framework in terms of the qubit number.
}

\keywords{Quantum Approximate Optimization Algorithm, denoising diffusion
probabilistic model, parameter initialization, Max-Cut}



\maketitle

\section{Introduction}\label{sec1}
Quantum computing is one of the major transformative technologies and presents an entirely new computational paradigm that has the potential to achieve an exponential or polynomial advantage for classically intractable problems encompassing machine learning \cite{gujju2024quantum,huang2021power,jerbi2023quantum,caro2022generalization}, molecular dynamics \cite{peruzzo2014variational,tilly2022variational,kandala2017hardware,kirby2021variational}, and combinatorial optimization \cite{farhi2014quantum,farhi2016quantum,cain2022qaoa,dupont2023quantum,alcazar2024enhancing}, etc. Currently, quantum computing is in the noisy intermediate-scale quantum (NISQ) era \cite{preskill2018quantum} where quantum hardware is characterized by limited qubit numbers, inherent system noise, significant quantum gate errors, and constrained qubit topologies. Despite these challenges, the community is dedicated to exploring quantum algorithms tailored to NISQ devices.  

Among these efforts, the Quantum Approximate Optimization Algorithm (QAOA) \cite{farhi2014quantum,farhi2016quantum,galda2021transferability,zhou2020quantum}, resorting to the hybrid quantum-classical mechanism and the use of the Parameterized Quantum Circuit (PQC), emerges as the foremost proposal. QAOA combines a parameterized quantum state evolution performed on a NISQ device with a classical optimizer used to find optimal parameters. QAOA has exhibited its effectiveness in addressing NP-Hard combinatorial optimization problems like Max-Cut \cite{farhi2014quantum,guerreschi2019qaoa,marwaha2021local,lu2023qaoa} via transforming the problems into the classical spin-glass model and then solving the corresponding ground-state energy. 
Further, the advancement of QAOA spurs the exploration of practical applications such as circuit layout designs \cite{jung2024qceda}, wireless communication \cite{liu2024quantum,zeng2024improved}, finance \cite{fellner2023parity,brandhofer2022benchmarking} and so on. 

Although QAOA shows immense practical significance in many fields, the parameter optimization, which is a crucial part of QAOA, presents a significant challenge \cite{zhou2020quantum,akshay2020reachability,bittel2021training} due to its non-convex cost landscape characterized by numerous low-quality and non-degenerate local minima hindering the trainability of the algorithm. Parameter optimization concentrates on evolving the initial parameters to the specific ones that can perform well on given tasks. To this end, several approaches have been proposed and can be categorized into two types, gradient-based and derivative-free methods \cite{wierichs2020avoiding,pellow2023qaoa,tibaldi2023bayesian}. However, these methods explicitly start the parameter optimization with random initialization, which often results in the difficulty of escaping a local minimum and then obtaining the low-performance solution. To overcome this challenge, a series of machine learning methods have been utilized to make significant breakthroughs in parameter initialization. Specifically, a meta-learning-based framework \cite{verdon2019learning} was proposed to facilitate the convergence of the parameter optimization in QAOA, where the propositional initial parameters are achieved by a recurrent neural network. In \cite{huang2022learning}, a similar idea was also investigated and extended to other variational quantum algorithm paradigms. 
Nevertheless, in
such cases they did not take into account the correlation between the pre-trained parameters and the corresponding
graph dataset. 
In this regard, Yuri et al. \cite{falla2024graph} explored the concentration effects of optimal parameters and applied graph representation learning technologies to predict good initial parameters through transferring pre-trained parameters. Despite some promising results obtained in this framework, the requirement for the similarity
metric computation between the target graph and all training graphs in the dataset results in high computational
complexity when the size of training graphs is large.

Diffusion models are a family of probabilistic generative models that progressively corrupt data by adding noise, and then learn to reverse this process for sample generation. Recent
research on diffusion models, especially denoising diffusion probabilistic
models (DDPMs) \cite{ho2020denoising},  score-based generative models (SGMs) \cite{nichol2021improved} and stochastic differential equations (Score SDEs) \cite{song2021maximum} have demonstrated significant advantages in image and visual generation. Currently, advanced tailored diffusion models, e.g., GLIDE \cite{nichol2021glide}, Imagen \cite{ruiz2023dreambooth} and DALL·E2 \cite{ramesh2022hierarchical}, have also shown the capacity to produce photorealistic images adopted by artists. Furthermore, the advancement in diffusion models has spurred the exploration of applications in other domains, such as quantum circuit compilation \cite{furrutter2024quantum}, neural architecture generation \cite{an2023diffusionnag}, and weights synthesizing for classical neural networks \cite{wang2024neural,soro2024diffusion}.

\begin{figure}[htbp]
    \centering
    \includegraphics[width= 0.80 \textwidth]{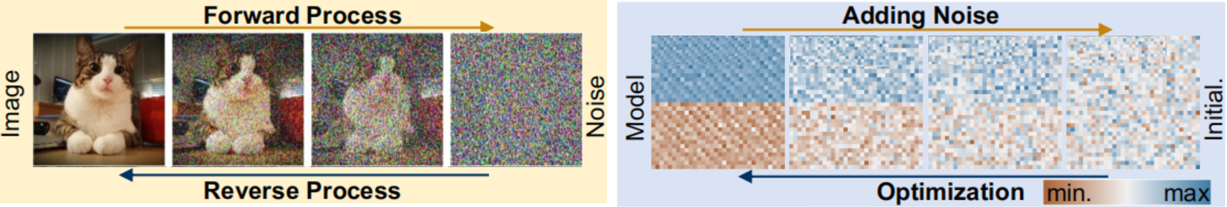} 
    \caption{ The left: illustrates the standard diffusion process in image generation. The right: denotes
the parameter distribution of QAOA.} \label{fig:1}
\end{figure}

Motivated by the work \cite{wang2024neural,soro2024diffusion}, taking a closer look at optimizing QAOA parameters and diffusion models, we are aware of the commonalities between diffusion-based image generation and QAOA parameter optimization process in the following aspects (illustrated in Figure \ref{fig:1}): $\left ( \romannumeral 1 \right ) $ \textit{both the reverse process of diffusion models and QAOA parameter optimization can be regarded as transitions from random noise/initialization to specific
distributions.} $\left ( \romannumeral 2 \right ) $ \textit{high-performing QAOA parameters and high-quality images can be degraded into simple distributions, e.g., Gaussian distributions, by iteratively adding noise.} To this end, inspired by the foregoing similarity, we delve into the synergy between the use of DDPM and QAOA parameter initialization where trained DDPM can explore the evolution from random parameters to high-performing initial parameters conditioned on the target graph. We posit that once DDPM is trained, it can efficiently learn the distribution of high-performing initial parameters conditioned on graph datasets and excel at generating a good initialization to improve the parameter optimization. This innovative approach paves the way for harnessing the capabilities of the generative machine learning model and quantum computing to address some of the most formidable challenges in computer science. 

The main contributions of our work are summarized as follows. 
\begin{itemize}
   \item \textbf{The graph representation learning} is performed adopting the typical methodology Graph2Vec \cite{narayanan2017graph2vec} for better graph conditioning since the latent representation of the graph can efficiently guide the generation of high-performing initial parameters.     
    \item  \textbf{DDPM} is adopted conditioning on the latent representation of the graph combining the pre-trained QAOA
parameters as input. The approach leverages the similarity between the DDPM-based image and QAOA initial
parameter generation, and the strengths of both quantum computing and the generative machine learning model.
 \item \textbf{Extensive evaluations} on various Max-Cut problem instances using
Xanadu Pennylane \cite{bergholm2018pennylane} show that the proposed scheme outperforms the random initialization strategy, indicating that, once trained, our model can efficiently generate high-quality initialization parameters for test instances, whose size can be even larger than those in the training dataset.
\end{itemize}

\section{Preliminaries}
\label{sec:Preliminaries}
\subsection{Quantum Computing}
Quantum mechanics operate within the Hilbert space $\mathcal{H} $, which is isomorphic to the complex Euclidean space $\mathbb{C}$. Dirac notation is used to denote quantum states, and a pure quantum state is defined by a column vector $\left | \cdot    \right \rangle$ (named `ket') with unit length. The mathematical expression of the $n$-qubit pure state is denoted as $\left | \psi  \right \rangle = {\textstyle \sum_{j=1}^{2^n}} \alpha _{j}\left | j  \right \rangle$ where $ {\textstyle \sum_{j=1}^{2^n}}   |\alpha _{j}|^{2}= 1$ and $\left | j  \right \rangle , j=1,2,\dots ,2^n$ stand for the computational basis states. 
In quantum computing, quantum gates, which can be represented by unitary matrices, are used to transform a quantum state to another without changing the norm.
The basic quantum gates can be classified into two groups, single-qubit and two-qubit gates. The commonly used single-qubit gates are the Pauli gates ($\sigma _{x}$, $\sigma _{y}$,  $\sigma _{z}$, and $I$) and their corresponding rotation gates, and the Hadamard gate ($H$), which can be represented by the following unitary matrices,

\begin{align}
\sigma _{x} =\begin{pmatrix}
  0& 1\\
  1&0
\end{pmatrix}
&& \sigma _{y} =\begin{pmatrix}
  0& -i\\
  i&0
\end{pmatrix}
&& \sigma _{z}=\begin{pmatrix}
  1&0 \\
  0&-1
\end{pmatrix}
&& I=\begin{pmatrix}
  1&0 \\
  0&1
\end{pmatrix}
&& H=\frac{1}{\sqrt{2} }\begin{pmatrix}
  1&1 \\
  1&-1
\end{pmatrix}
\end{align}
\begin{align}
      R_{x} \left ( \theta  \right ) =\begin{pmatrix}
 \cos\frac{\theta}{2}  & -i\sin\frac{\theta}{2} \\
  -i\sin\frac{\theta}{2} & \cos\frac{\theta}{2}
\end{pmatrix}
 &&   R_{y} \left ( \theta  \right ) =\begin{pmatrix}
 \cos\frac{\theta}{2}  & -\sin\frac{\theta}{2} \\
  \sin\frac{\theta}{2} & \cos\frac{\theta}{2}
\end{pmatrix}
&& R_{z} \left ( \theta  \right ) =\begin{pmatrix}
 e^{-i\frac{\theta}{2}}  & 0 \\
  0 & e^{i\frac{\theta}{2}}
\end{pmatrix}
\end{align}
The notable two-qubit gate is the CNOT gate shown as
\begin{align}
  CNOT=\begin{pmatrix}
 1 & 0 & 0 & 0\\
 0 & 1 & 0 & 0\\
 0 & 0 & 0 & 1\\
 0 & 0 & 1 &0
\end{pmatrix}
\end{align}

\subsection{Quantum Approximate Optimization Algorithm}
\begin{figure}[htbp]
    \centering
    \includegraphics[width= 0.9 \textwidth]{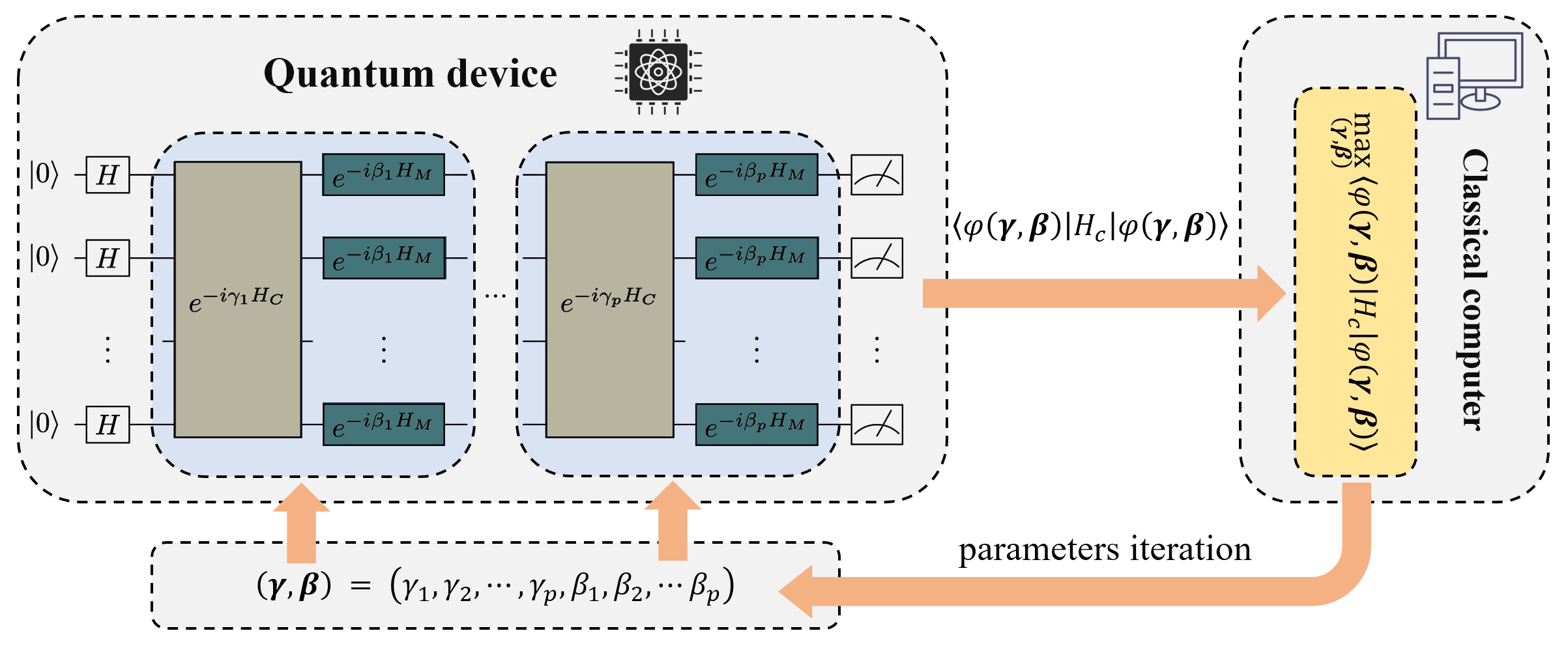} 
    \caption{ The overall workflow of QAOA.} \label{fig:2}
\end{figure}

The Quantum Approximate Optimization Algorithm (QAOA) is a paradigm of variational quantum algorithms developed to address combinatorial optimization tasks on NISQ devices. It utilizes the problem-dependent parameterized quantum circuit involving alternate Hamiltonian and Mixer layers
to prepare and evolve quantum states that encode potential solutions to the optimization problem. By adjusting the variational parameters using classical optimization techniques, QAOA aims to find near-optimal solutions efficiently, as illustrated in Figure \ref{fig:2}. QAOA is particularly suited for the Max-Cut problem and leverages the expressive power of quantum computation to explore large solution spaces more effectively than its classical counterparts. Its efficacy lies in classical initialization and parameter optimization with the aim of exploring the parameter space to achieve good approximation solutions.


In QAOA, the binary assignment combinatorial optimization is first encoded in a cost Hamiltonian $H_{c}$ by the mapping between $n$ classical binary decision variables $s_{i}\in \left \{ 0,1 \right \} , i=1,2,\dots ,n$ and the eigenvalues of the quantum Pauli $\sigma _{z}$, where the ground energy eigenstate of $H_{c}$ corresponds to the solution of the combinatorial optimization problem. Second, the transverse field mixer Hamiltonian is constructed as $H_{m} =  {\textstyle \sum_{i=1}^{n}} \sigma _{x}^{i}$. Then, the initial state of uniform superposition $\left | +    \right \rangle ^{\otimes n}$ is prepared by performing Hadamard gates on all qubits with an all-zero state. Next, a variational quantum state $\left | \psi \left ( \gamma _{1},\dots ,\gamma _{p} ,\beta _{1},\dots ,\beta _{p}    \right )  \right \rangle $ is prepared by employing alternate Hamiltonian and Mixer layers $e^{-i\gamma _{i} H_{c} } $ and $e^{-i\beta _{i} H_{m} }$, $i=1,2,\dots ,p$ as follows
\begin{equation}
     \left | \psi \left ( \gamma _{1},\dots ,\gamma _{p} ,\beta _{1},\dots ,\beta _{p}    \right )  \right \rangle=e^{-i\beta_{p}H_{m}}e^{-i\gamma _{p}H_{c}}\cdots e^{-i\beta_{1}H_{m}}e^{-i\gamma _{1}H_{c}}\left | +    \right \rangle ^{\otimes n}   
\end{equation}
Finally, the selected classical optimizer is employed to vary parameters $\gamma_{i}$ and $\beta_{i}$, $i=1,2,\dots ,p$ to minimize the cost function 
\begin{equation}
  C\left ( \gamma _{1},\dots , \gamma _{p}, \beta _{1},,\dots ,\beta _{p} \right )   =\left \langle \psi \left ( \gamma _{1},\dots ,\gamma _{p} ,\beta _{1},\dots ,\beta _{p}    \right ) \right| H_{c}\left | \psi \left ( \gamma _{1},\dots ,\gamma _{p} ,\beta _{1},\dots ,\beta _{p}    \right )  \right \rangle 
\end{equation}

For the Max-Cut problem, given a graph $G=\left(V,E\right)$ where $V$ is the set of nodes and $E$ is the set of edges, the goal of the Max-Cut is to partition the
set of nodes $V$ into two disjoint subsets such that the total
weight of the edges connecting the two subsets is maximized. The mathematical expression of the Max-Cut is formulated as follows
\begin{equation}
\underset{\vec{s} }{\min }  \sum_{ \left ( i,j \right )\in E }^{} -w_{ij}s_{i}s_{j} 
\end{equation}
where $s_{k} \in \left \{ -1,1 \right \}  $ and the weight $w_{ij}$ of an edge $(i,j)$ is set to $1$ for the unweighted Max-Cut problems. To apply QAOA to the Max-Cut, the above objective is encoded into the following problem Hamiltonian by mapping binary variables $s_{k}$ onto the eigenvalues of the Pauli $\sigma _{z}$
\begin{equation}
     \sum_{\left ( i,j \right )\in E }^{} -w_{ij}\sigma _{z}^{i}\sigma _{z}^{j}
\end{equation}
Thus, minimizing the objective of the Max-Cut is equivalently transformed into obtaining the ground-state energy of the problem Hamiltonian.

\subsection{Graph Representation Learning}
Graph representation learning focuses on learning the latent representation of nodes, edges, and entire graphs that can be used for various downstream tasks. The goal is to transform the input graph data into a low-dimensional vector space while preserving the structural and semantic properties of the graph. In this domain, several approaches have been proposed, such as Laplacian Eigenmaps \cite{belkin2001laplacian}, Node2Vec \cite{grover2016node2vec}, Graph2Vec \cite{narayanan2017graph2vec} and DeepWalk \cite{perozzi2014deepwalk}. Among these methods, Graph2Vec is appealing to us, since it is capable of capturing local properties of the graph, such as the type of subgraphs as well as the parity of nodes \cite{falla2024graph}, closely related to optimal QAOA parameter concentration effects which is beneficial to distill knowledge from pre-trained optimal parameters. 
Graph2Vec first employs the Weisfeiler-Lehman kernel \cite{shervashidze2011weisfeiler} to extract the features of the rooted subgraph and then feeds them into a Doc2Vec model \cite{lau2016empirical} to obtain the final embeddings. Therefore, in this paper, we use Graph2Vec to embed the latent representation of the graph dataset.

\subsection{Denoising Diffusion Probabilistic Model}
Denoising diffusion probabilistic model (DDPM) \cite{ho2020denoising} refines the diffusion model with a training paradigm characterized by forward and reverse processes in multi-step chains indexed by
time steps. The forward process, dubbed the diffusion process, is characterized as progressively perturbing data to noise, while the backward process, referred to as the denoising process, gradually converts noise back to data. The goal of the former is to transform any distribution into a simple prior, while the latter is to reverse the former by learning a transition kernel that utilizes deep neural networks.

\textbf{Diffusion process.} Given an original sample $x_{0}$, Gaussian noise is progressively added for $T$ steps to obtain a series of noisy samples $x_{1}, x_{2}, \dots, x_{T}$. Based on the reparameterization rule, the mathematical expression of this process at any time step $t$ can be written as follows \cite{ho2020denoising},
\begin{equation}
    x_{t} = \sqrt{\bar{\alpha}  _{t} } x_{0} + \sqrt{1-\bar{\alpha}  _{t}} \varepsilon  \quad  or \quad q\left ( x_{t}|x_{0} \right )=\mathcal{N  }\left ( x_{t}; \sqrt{\bar{\alpha}  _{t} } x_{0}, \left ( 1-\bar{\alpha}  _{t} \right )\varepsilon ^2 \right )   
\end{equation}
where $\bar{\alpha}  _{t}$ is a constant decreasing with $t$, and $\mathcal{N}$ denotes the Gaussian distribution, $\varepsilon$ is the sample from Gaussian noise.

\textbf{Denoising process.} Beginning with the time step $T$, taking a sample $x_{T}$ and the current time step $T$ as inputs, the noise evaluated by a deep neural network is removed from $x_{T}$ to generate a new sample $x_{T-1}$.
The denoising process repeats this process until the new data sample is restored. Mathematically, the denoising process can be formulated as follows \cite{ho2020denoising}
\begin{equation}
    \mathbf{x}_{t-1}\sim \frac{1}{\sqrt{\alpha_{t}}}\left(\mathbf{x}_{t}-\frac{1-\alpha_{t}}{\sqrt{1-\bar{\alpha}_{t}}} \boldsymbol{\varepsilon}_{\theta}\left(\mathbf{x}_{t}, t\right)\right)
\end{equation}
where $\alpha_{i}$ is a hyperparameter chosen ahead of model training and $\boldsymbol{\varepsilon}_{\theta}\left(\mathbf{x}_{t}, t\right)$ is a deep neural network to estimate the noise at the current time step. Specially, $\boldsymbol{\varepsilon}_{\theta}\left(\mathbf{x}_{t}, t\right)$ is trained by minimizing the following loss function
\begin{equation}
    \theta\leftarrow \theta-\nabla_{\theta}\left\|\boldsymbol{\varepsilon}-\boldsymbol{\varepsilon}_{\theta}\left(\sqrt{\bar{\alpha}_{t}} \mathbf{x}_{0}+\sqrt{1-\bar{\alpha}_{t}} \boldsymbol{\varepsilon}, t\right)\right\|^{2} 
\end{equation}

\section{Parameter Generation for QAOA with Conditional DDPM}
\label{sec:others}
\textbf{Problem Definition:} In the proposed work, we devote ourselves to the following problem: Given a set $W_{\theta } =\left \{ \mathbf{\vec{\theta}  }_{1}, \mathbf{\vec{\theta}  }_{2} ,\cdots ,\mathbf{\vec{\theta}  }_{M}   \right \} $ of pre-trained QAOA parameters for $M$ different Max-Cut problem instances $\left \{ G_{1}, G_{2} ,\cdots ,G_{M}  \right \}$, DDPM intends to learn the distribution $p\left(W_{\theta}\right)$ of the pre-trained parameters of problem instances such that for a new graph $G_{new}$, we can conditionally generate the high-performing initial parameters with $p\left ( \mathbf{\vec{\theta}}  _{new}\mid G _{new} \right )$, which is capable of achieving good
performance for the new instance. Intuitively, given the pre-trained parameter set, DDPM can construct a clear relationship between the optimized QAOA parameters and the corresponding graph data. For an unseen graph, we can compute the graph representation and use this to guide DDPM in generating parameters that mimic trained parameters. 

\begin{figure}[htbp]
    \centering
    \includegraphics[width= 1.0 \textwidth]{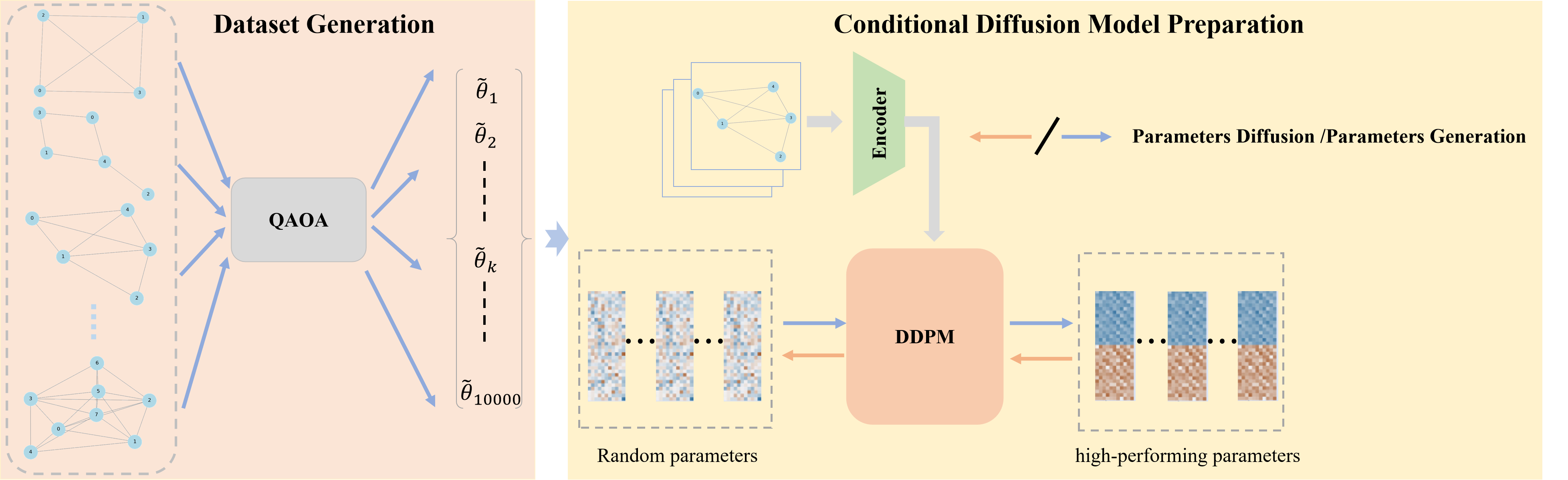}
    \caption{The flow chart of conditional DDPM based parameter generation } \label{fig:3.5}
\end{figure}

In this section, we propose an initial parameter diffusion framework for QAOA, which aims to generate high-performing initial parameters from random noise. As illustrated in Figure \ref{fig:3.5}, our approach consists of two processes, named dataset generation, and conditional DDPM preparation consisting of parameter diffusion and generation.
Due to the lack of publicly available data sets, we first construct the dataset consisting of the graph dataset and the correspondingly pre-trained parameters of QAOA for the Max-Cut problems. Subsequently, we detail the initial parameter generation based on conditional DDPM for QAOA. The model is trained to synthesize high-performing initial parameters from random noise conditioned on the graph dataset via the following chain: random noise $\rightarrow$ denoising process $\rightarrow$ generated initial parameters. 

\subsection{Dataset Generation for Conditional DDPM}
\label{dataset}
\begin{figure}[htbp]
\label{sec:data}
    \centering
    \includegraphics[width= 0.80 \textwidth]{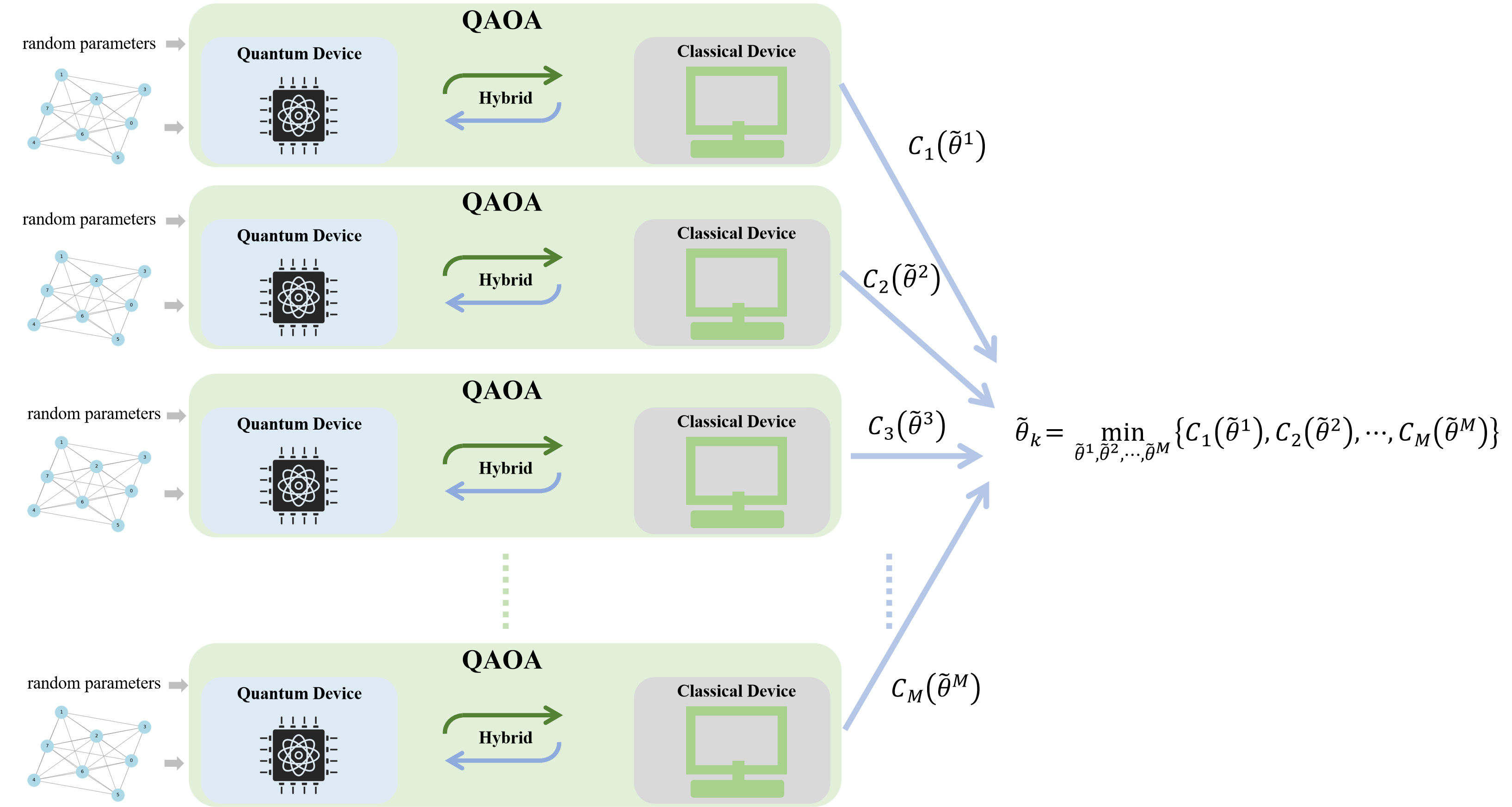} 
    \caption{The flow of the multi-start strategy for QAOA parameter optimization} \label{fig:4.5}
\end{figure}

We first construct a synthetic graph dataset $\mathcal{G} =\left \{ G_{1}, G_{2},\dots ,G_{10000} \right \}$ comprising 10000 instances. The graph dataset is produced with the following rules: we start with 3500 random graphs with node sizes $N$ ranging from $4$ to $8$, and the probability $p$ of having an edge between two nodes varying between $0.3$ and $0.75$. $4000$ regular graphs are constructed with node sizes ranging from $4$ to $8$ and the degree of each node ranging from $3$ to $7$. Cautiously, the product of the number of nodes and the degree of each node is restricted to even numbers. 2500 Watts-Strogatz graphs are produced with node sizes between $4$ and $8$. Moreover, we set the number of neighbor nodes of each node and the probability of reconnection to be $3$ and range from $0.3$ to $0.75$, respectively. Next, each of the above graph dataset is inputted into QAOA for the Max-cut problem. To maintain consistency between the dimensions of the input and output of the diffusion model, we fixed the layers of QAOA to be $3$, which contains $6$ parameters $\mathbf{\vec{\gamma}  } =\left (\gamma_{1}, \gamma_{2}, \gamma_{3}  \right ) $ and $ \mathbf{\vec{\beta}  } =\left (\beta_{1}, \beta_{2}, \beta_{3} \right )$. Subsequently, we randomly initialize parameters $\mathbf{\vec{\gamma}  }$ and $\mathbf{\vec{\beta}  }$, and
recursively perform the parameter optimization process using the SGD optimizer \cite{ruder2016overview}  for $500$ iterations. Since the inherently complex optimization landscape of QAOA \cite{zhou2020quantum, akshay2020reachability}, the final optimized parameters may have a relatively large gap compared to the optimal ones. Therefore, as characterized in Figure \ref{fig:4.5}, we devise a multi-start strategy to find the parameter setting that is closer to the optimal one. In the proposed scheme, for each Max-cut instance, multiple QAOAs are parallelly optimized from scratch, starting with multiple random initial parameters.
The optimized parameters corresponding to the lowest cost of the problem are flattened into a vector and recorded as $\mathbf{\vec{\theta}  }_{k}\in R^{6}$. The saved sets of parameters $S=\left [ \mathbf{\vec{\theta}  }_{1}, \mathbf{\vec{\theta}  }_{2},\dots ,\mathbf{\vec{\theta}  }_{k},\dots ,\mathbf{\vec{\theta}  }_{10000}\right ] $ and graphs $\mathcal{G} =\left \{ G_{1}, G_{2},\dots ,G_{10000} \right \}$ are utilized to train the diffusion model.   

\subsection{Conditional Diffusion Model Training and Inference}

\begin{figure}[htbp]
    \centering
    \includegraphics[width= 0.98 \textwidth]{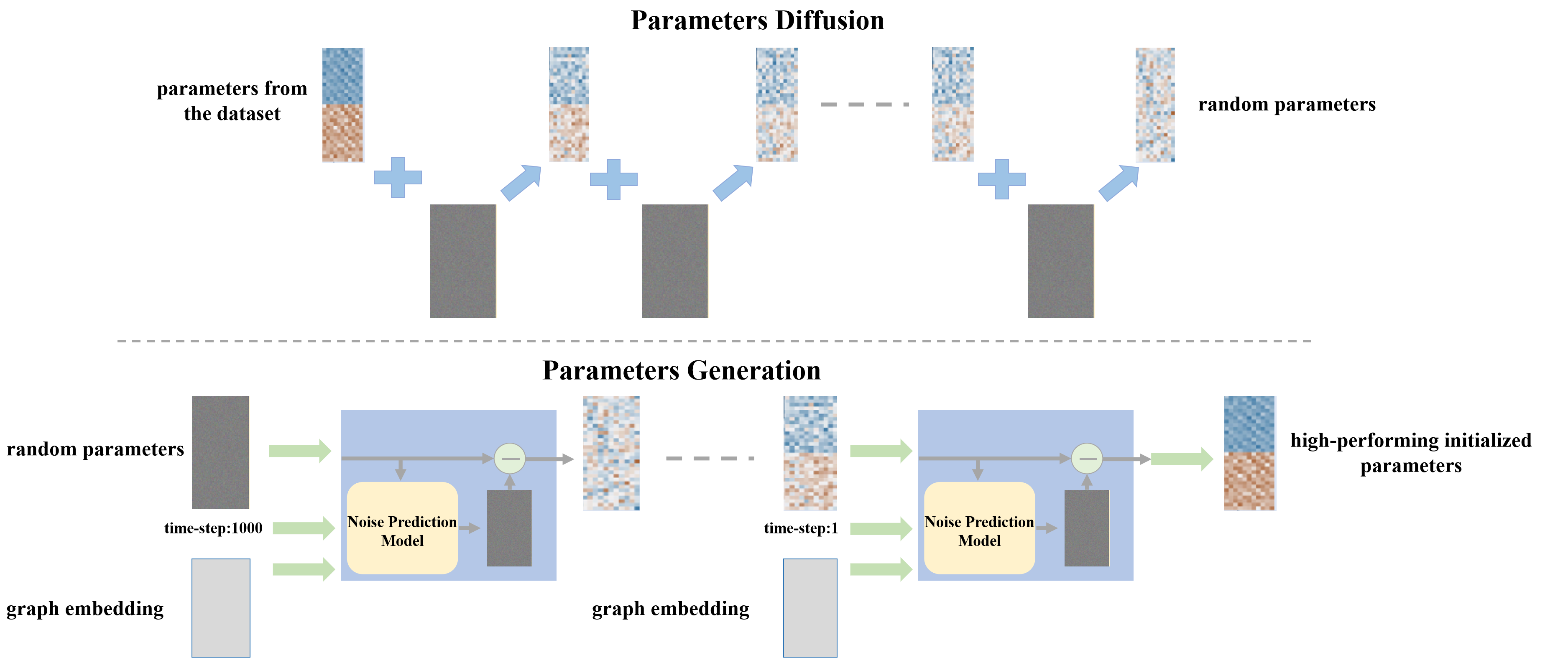}
    \caption{ 
    Illustrated on the top is the parameter diffusion process, and on the bottom is the parameter generation process.} \label{fig:4}
\end{figure}

In pursuit of enhancing the application of QAOA to the Max-Cut problem, the proposed approach involves the integration of the conditional DDPM model and QAOA parameter initialization. We first reshape the pre-trained parameters in $S$ into the tensor representation and transform each graph in $\mathcal{G}$ into their correspondingly latent representation.  
Next, as illustrated in Figure \ref{fig:4}, the model is built with two processes, parameter diffusion and parameter generation. In the parameter diffusion process, we progressively corrupt the pre-trained parameters in the training dataset with Gaussian noise where the neural network is trained by taking current noisy parameters, the latent representation of the graph, and time step as input to predict the current noise. The parameter generation process inputs random noise into the trained neural network conditioned on the latent representation of the target graph to infer the high-performing initial parameters for QAOA. 

\subsubsection{Parameter Diffusion Process}
We first take the pre-trained parameter sample from the training dataset prepared in the last subsection \ref{dataset} and add the Gaussian noise to it. Concomitantly, the neural network, dubbed noise prediction network, takes the noisy parameter sample, the correspondingly latent representation of the graph and the current time step as inputs and learns to predict the noise of the present step, aiming at subtracting it from the input sample. Specifically, slightly modifying the work \cite{ho2020denoising}, the parameter diffusion process can be characterized in Algorithm \ref{DDPMTra}, where $x_{0}$ and $z^{g}_{\mathcal{D} }$ depict the sample of noiseless parameters and the latent representation of the graph in the training dataset, respectively, $\varepsilon$ is the added Gaussian noise, $T$ is the total time step and $\mathcal{N}(\mathbf{0}, \mathbf{I})$ represents the Gaussian noise distribution. We input the noisy parameter sample $\sqrt{\bar{\alpha}_{t}} \mathbf{x}_{0}+\sqrt{1-\bar{\alpha}_{t}} \boldsymbol{\varepsilon}$, the latent representation $z^{g}_{\mathcal{D} }$ and the current time step $t$ into the noise prediction network $\boldsymbol{\varepsilon}_{\theta}\left(\cdot \right)$ parameterized by $\theta$ to approximate the present noise $\varepsilon$.

\begin{algorithm}
\caption{The Noise Prediction Network Training}\label{DDPMTra}
\begin{algorithmic}[1]

\State repeat
        \State \quad  $\mathbf{x}_{0} \sim S$
		\State \quad  $g \sim \mathcal{G}$
		\State \quad  $z^{g}_{\mathcal{D} } \longleftarrow encoder \left(g\right)$ via Graph2Vec
        \State \quad  $t \sim \operatorname{Uniform}(\{1, \ldots, T\})$
        \State \quad  $\varepsilon \sim \mathcal{N}(\mathbf{0}, \mathbf{I})$
        \State \quad Take gradient descent step on $\nabla_{\theta}\left\|\boldsymbol{\varepsilon}-\boldsymbol{\varepsilon}_{\theta}\left(\sqrt{\bar{\alpha}_{t}} \mathbf{x}_{0}+\sqrt{1-\bar{\alpha}_{t}} \boldsymbol{\varepsilon},z^{g}_{\mathcal{D} }, t\right)\right\|^{2}$
        \State until converged
\end{algorithmic}
\end{algorithm}

\subsubsection{parameter Generation Process}
After finishing the training of the noise prediction network in parameter diffusion, we feed random parameters and target graph representation into the trained network to generate high-performing initial parameters.
The detailed parameter generation or inference process is characterized in Algorithm \ref{Sampling}, where $x_{T}$ is the random parameter, $z^{G}_{\mathcal{D} }$ is the latent representation of the target graph, $\mathcal{N}(\mathbf{0}, \mathbf{I})$ represents the Gaussian noise distribution and $\sigma_{t}$ is a constant variance. The reverse process in step $5$ is utilized to eliminate the noise from the current sample, aiming at ultimately outputting the high-performing parameter $x_{0}$ for QAOA.

\begin{algorithm}
\caption{parameter Generation} \label{Sampling}
\begin{algorithmic}[1]
\State $\mathbf{x}_{T} \sim \mathcal{N}(\mathbf{0}, \mathbf{I})$
\State $z^{G}_{\mathcal{D} } \longleftarrow encoder \left(G\right)$ via Graph2Vec, where $G$ is the target graph
		\State \text { for } $t=T, \ldots, 1 \text { do }$
		\State \quad \quad $\mathbf{z} \sim \mathcal{N}(\mathbf{0}, \mathbf{I})$ \text { if } $t>1$ \text {, else } $\mathbf{z}=\mathbf{0}$
        \State \quad \quad $\mathbf{x}_{t-1}=\frac{1}{\sqrt{\alpha_{t}}}\left(\mathbf{x}_{t}-\frac{1-\alpha_{t}}{\sqrt{1-\bar{\alpha}_{t}}} \boldsymbol{\varepsilon}_{\theta}\left(\mathbf{x}_{t},z^{G}_{\mathcal{D} }, t\right)\right)+\sigma_{t} \mathbf{z}$
        \State end for
        \State return $x_{0}$
\end{algorithmic}
\end{algorithm}

\section{Experiment}
In this section, we conduct experiments to validate the effectiveness of the proposed framework.
\subsection{Implementation Detail}
\begin{figure}[htbp]
\label{sec:model}
    \centering
    \includegraphics[width= 0.6 \textwidth]{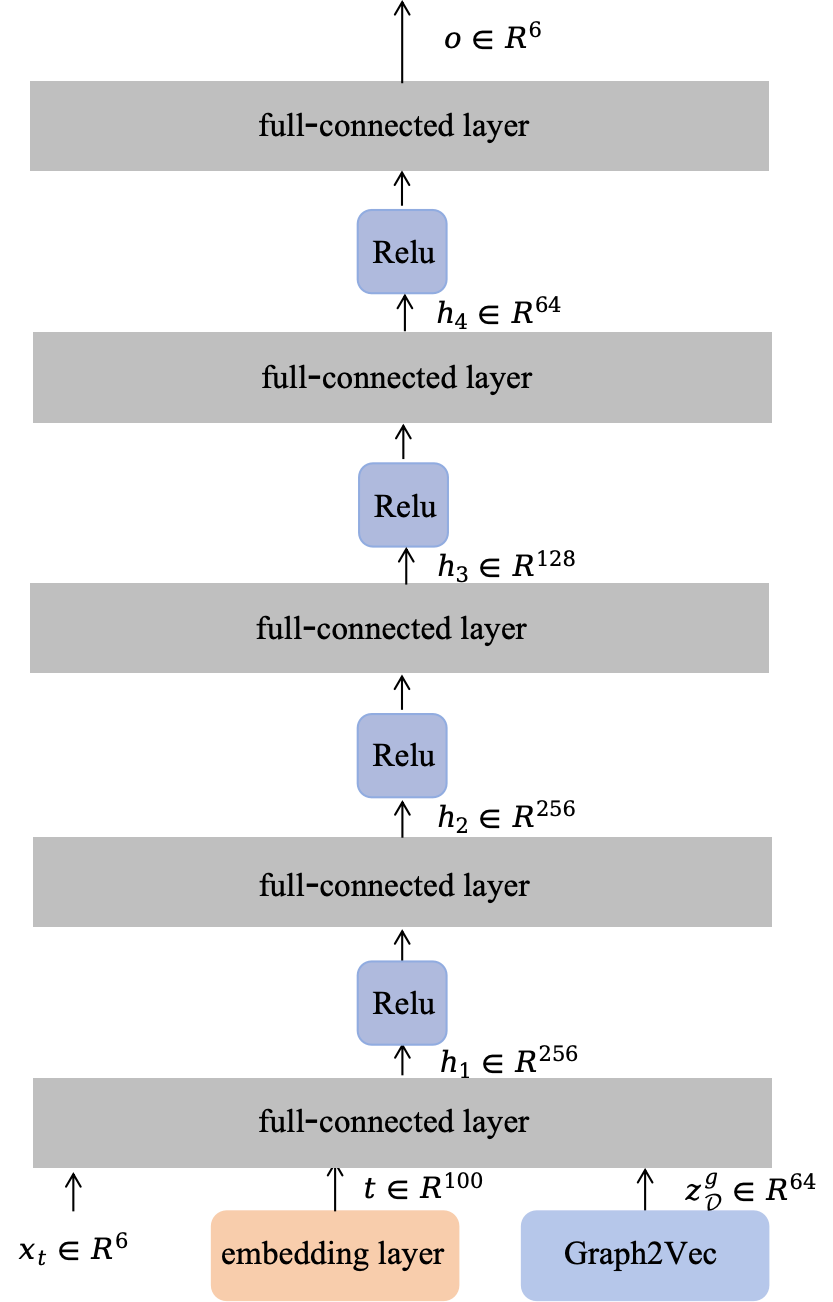} 
    \caption{ The architecture of the noise prediction network.} \label{fig:5}
\end{figure}

In the proposed framework, the noise prediction network is constructed as illustrated in Figure \ref{fig:5}. The input parameters, the time step encoding by the embedding layer and the graph representation by the Graph2Vec are fed into a $5$ full-connected layers
with ReLU nonlinearity to predict the noise of the present step. In particular, we set the dimension of the embedding layer to be $100$. For the Graph2Vec model, we use the graph-embedding module from the karateclub \cite{rozemberczki2003api}, an unsupervised machine learning package,  which is also an extension to NetworkX. The training is performed with $64$ embedding dimensions and $100$ epochs. Moreover, we set the down-sampling rate, the number of workers and the learning rate to be $ 0.0001$, $4$ and $0.065$, respectively. 
The noise prediction network is trained for 100 epochs with batch size being $50$. Moreover, we use the Adam optimizer to optimize the model with the learning rate being $10^{-3}$, and neither the weighted decay nor the learning rate scheduler is used.

\subsection{Dataset and Baseline Model}
We first introduce the approximation ratio metric defined as $r^{\star }=\frac{C_{QAOA}}{C^{\star }}$, where $C_{QAOA}$ and $C^{\star }$ are the solutions of QAOA and the brute-force search approach, respectively, and set the random initialization strategy as the baseline. Subsequently, we show the validation of Graph2Vec and the advantage of the proposed method on several Max-Cut tasks in and beyond the training regime.

\subsubsection{Graph2Vec Setup}
\begin{figure}[htbp]
    \centering
    \includegraphics[width= 0.56 \textwidth]{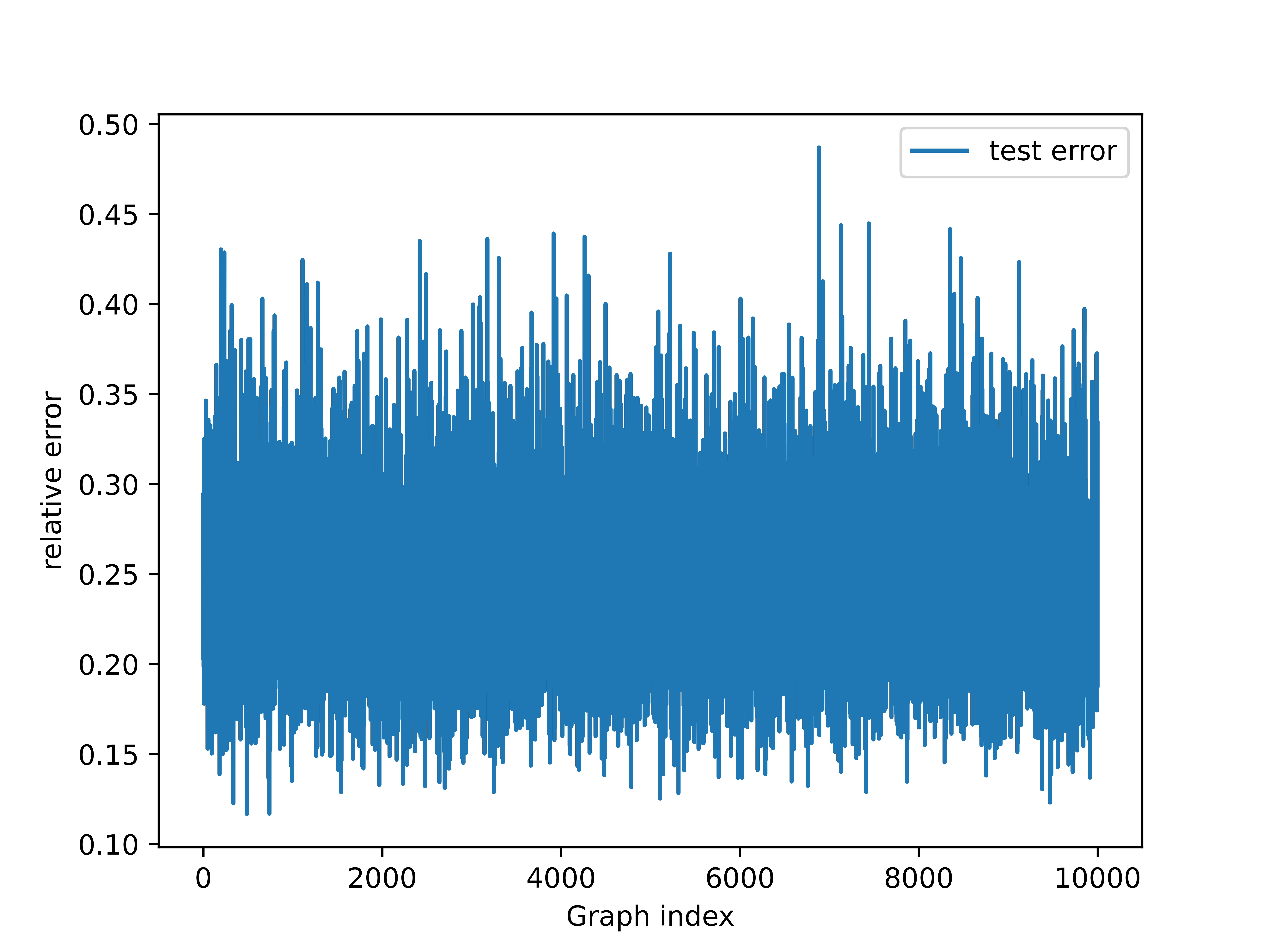} 
    \caption{The relative error of graph dataset representation learning via the Graph2Vec model.} \label{fig:g2v}
\end{figure}

To validate the availability of Graph2Vec, the graph dataset in the training dataset is reused as a test set, namely, projecting the graph dataset again into the embedding space. We figure up the relative error between the feature vectors of the training and the test sets, as illustrated in Figure \ref{fig:g2v}. The result shows that all relative errors are less than 0.5 with the embedding dimension being 64. 

\subsubsection{Task Setup}
\begin{figure}[htbp]
    \centering
    \includegraphics[width= 0.56 \textwidth]{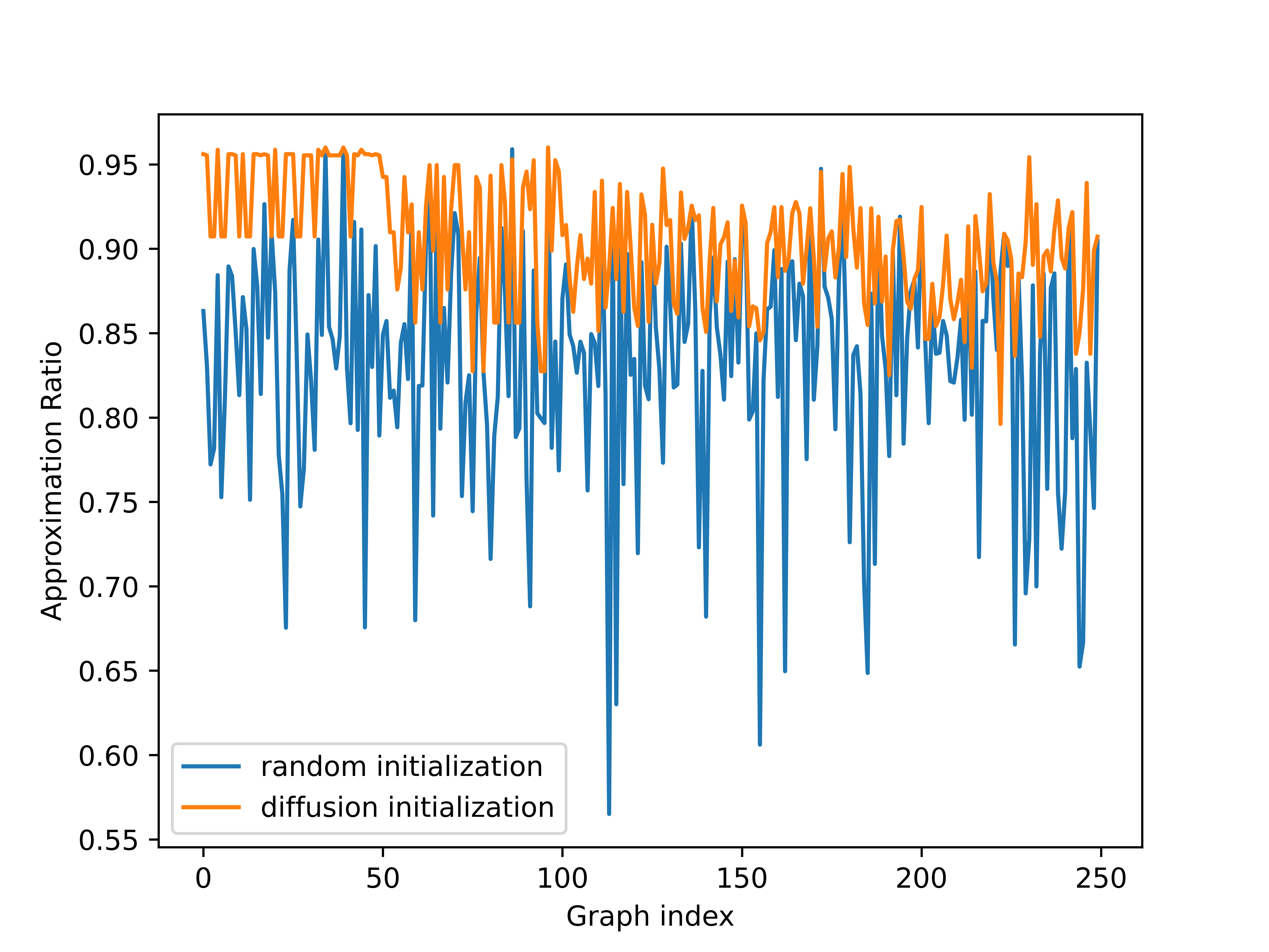} 
    \caption{ Comparison of the approximation ratio between the diffusion-based parameter initialization and the baseline for the random graph dataset.} \label{fig:6}
\end{figure}

First, $250$ random test graphs for the Max-Cut problem are randomly generated to demonstrate the improvement of the approximation ratio derived by the conditional DDPM-based parameter initialization over the baseline. We proceed to iteratively optimize QAOA's parameters for 100 steps starting with the proposed and random initialization strategies. The results are presented in Figure~\ref{fig:6}, where the orange line is the approximation ratio of the proposed
initialization strategy, and the blue line represents the random initialization. The average improvements for different node sizes and edge probabilities compared to the baseline are listed in Table \ref{table_random}. The results show that the proposed framework always outperforms the baseline and achieves a higher approximation ratio, up to $14.4\%$, and $7.49\%$ improvement on average. 

\begin{figure}[htbp]
    \centering
    \includegraphics[width= 0.56 \textwidth]{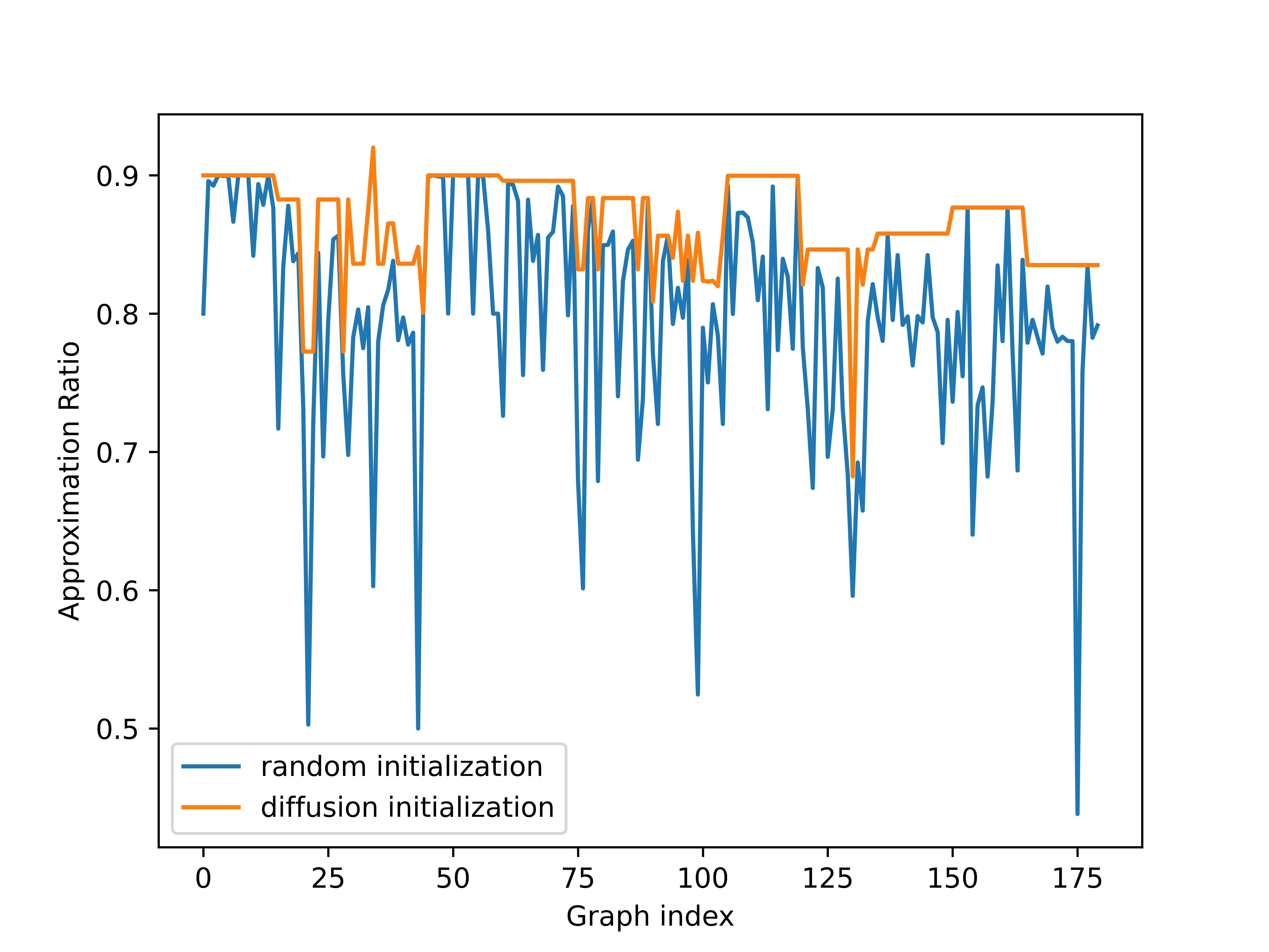} 
    \caption{Comparison of the approximation ratio between the diffusion-based parameter initialization and the baseline for regular graph dataset.} \label{fig:regular}
\end{figure}

Then, we construct the regular graph set consisting of $180$ graphs for the Max-Cut problem with the degree from $3$ to $7$ and node sizes varying between $4$ and $8$. For this dataset, QAOA starts with the proposed and random initialization strategies and continues to optimize parameters for 100 steps. As illustrated in Figure~\ref{fig:regular}, the result demonstrates that our method always outperforms the baseline and achieves a higher approximation ratio, up to $11.0\%$, and $8.31\%$ improvement on average. The average improvements for different node sizes and degrees compared to the baseline are listed in Table \ref{sec:table_r}.

\begin{figure}[htbp]
    \centering
    \includegraphics[width= 0.56 \textwidth]{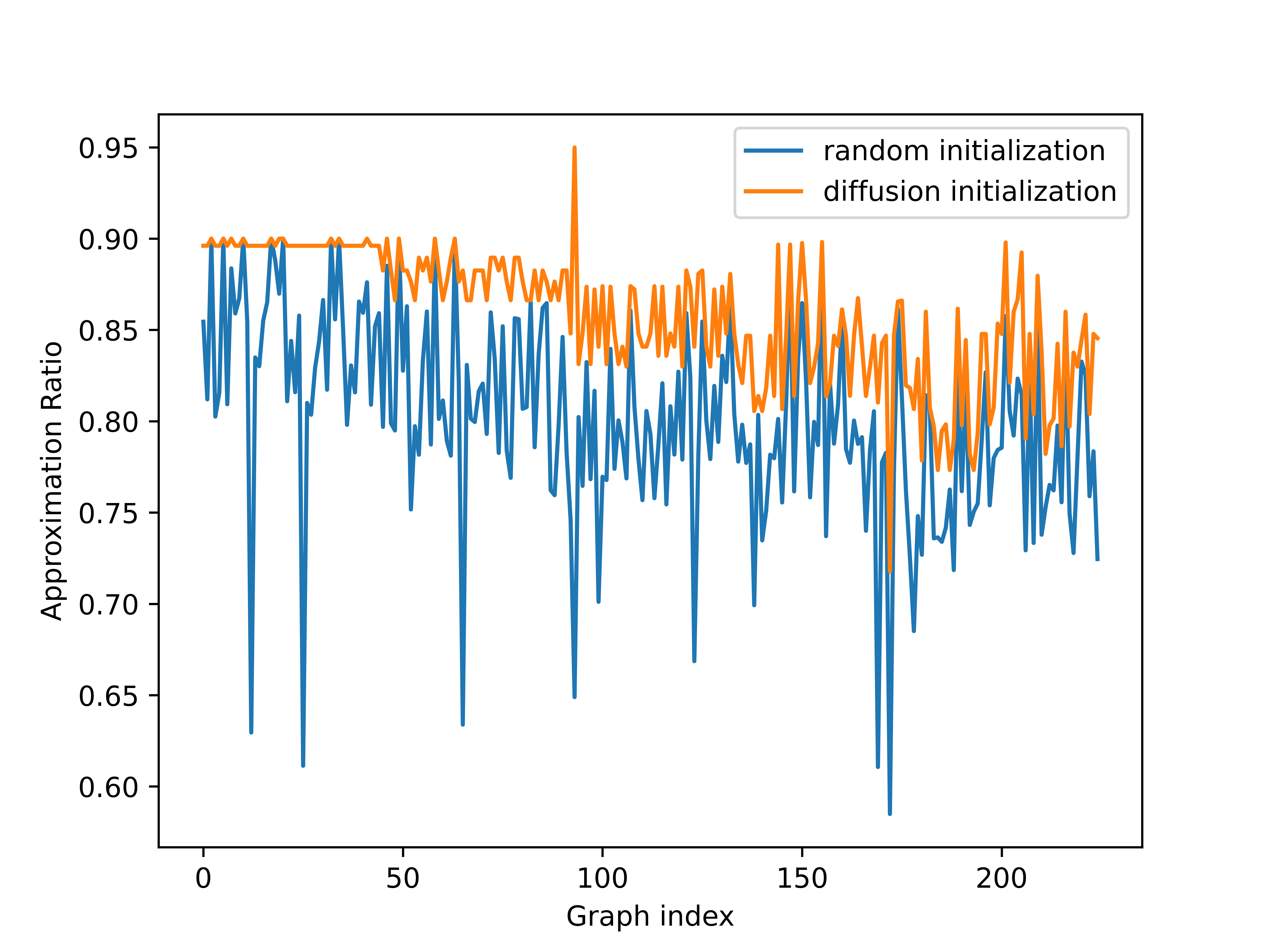} 
    \caption{Comparison of the approximation ratio between the diffusion-based parameter initialization and the baseline for the Watts-Strogatz graph dataset.} \label{fig:Watts}
\end{figure}

Next, $225$ Watts-Strogatz graphs with the number of nearest neighbors being fixed to $3$ are generated to show the advantage of our proposed initialization strategy over random initialization. Analogously to the previous experiment, these initial parameters are also fed to the QAOA, the optimization step of which is also set to 100, for further optimization. The results shown in Figure~\ref{fig:Watts} and Table \ref{tab:Watts} demonstrate that the proposed method can obtain an approximation ratio improvement of $6.08\%$ on average and up to $11.4\%$. 

\begin{figure}[htbp]
    \centering
    \includegraphics[width= 0.75 \textwidth]{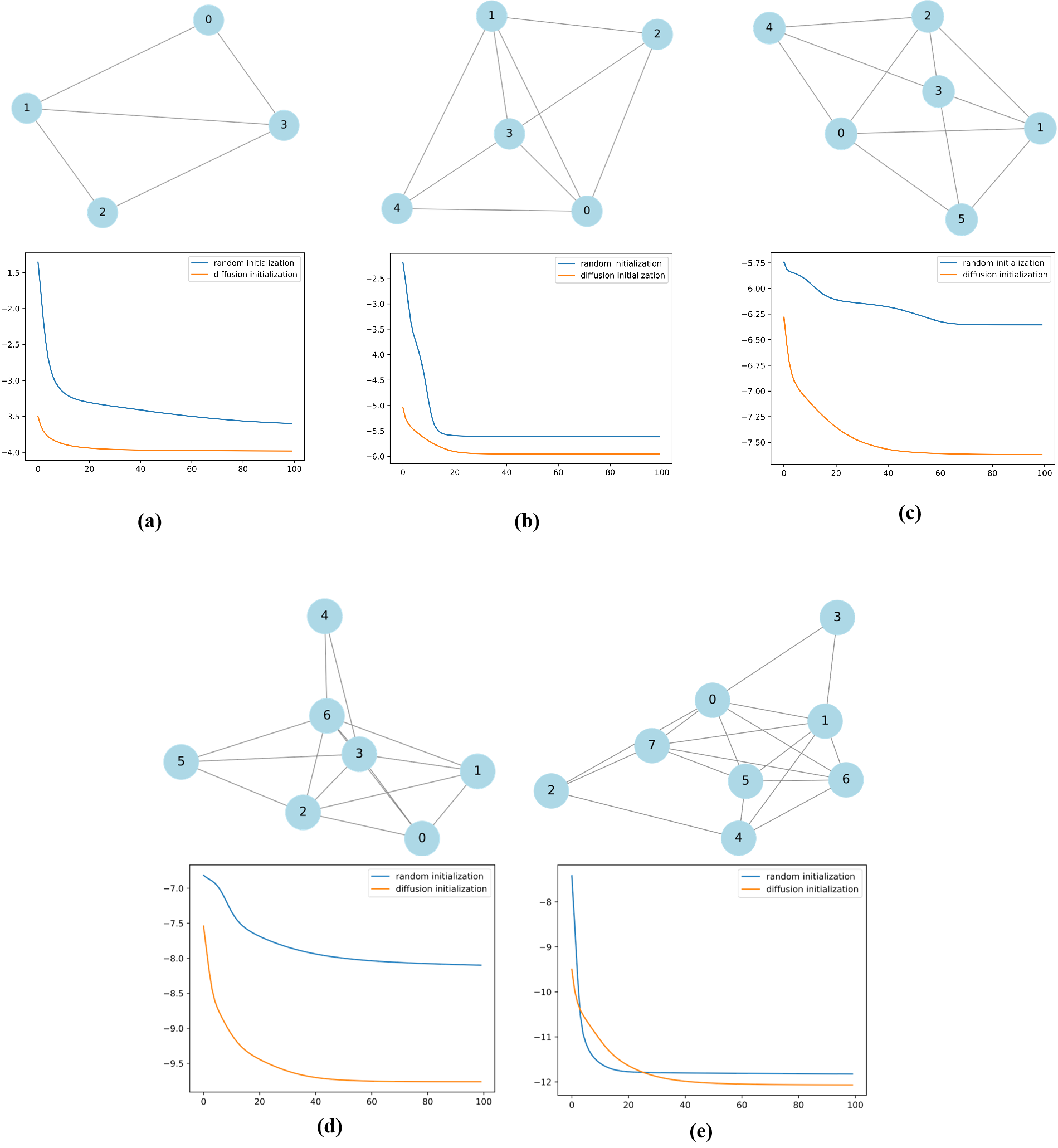} 
    \caption{ Cost convergence of the proposed method and the baseline.} \label{fig:7}
\end{figure}

Moreover, for clarity, we randomly pick 5 graphs with different node sizes from the above graph datasets to show the convergence of the proposed work against the baseline, as illustrated in Figure \ref{fig:7}. Based on the results, we discover that the scheme can render QAOA converging to lower energy/cost compared to the baseline. 

\begin{figure}[htbp]
    \centering
    \includegraphics[width= 1.0 \textwidth]{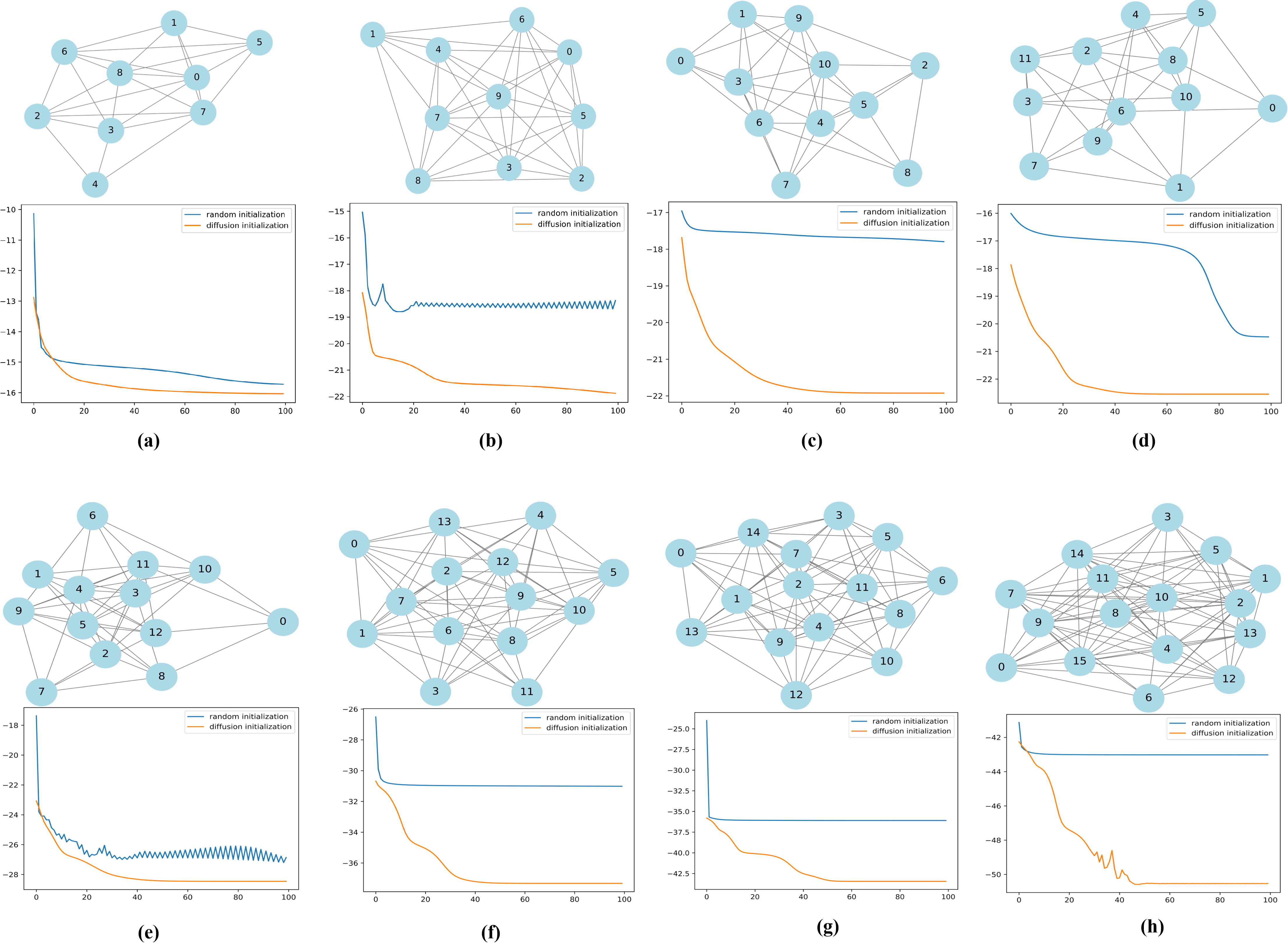} 
    \caption{ Comparison of the diffusion-based parameter initialization and the baseline for larger instances.} \label{fig:8}
\end{figure}

Furthermore, to validate the availability of the proposed approach for being extrapolated to larger instances, 
we first generate sets of random, regular and Watts-Strogatz graphs. For each type of graph, we generate $400$ instances with node sizes varying from 9 to 16 (50 instances for each node size).
Next, for each graph, we perform QAOA with optimization steps being 100, and the
average improvements for different node sizes compared to the baseline are listed in Table \ref{tab:large}. The results show that our approach achieves a higher approximation ratio, up to $28.4\%$, and a $12.1\%$ improvement over the baseline, demonstrating that the advantage of our approach can be extrapolated outside the training dataset regime. 
Moreover, as shown in Figure \ref{fig:8}, we reveal that the proposed framework is capable of rendering QAOA
converging to lower energy/cost over the baseline, where $8$ graphs are randomly picked from the above $1200$ graphs.

In addition to the performance comparison, we assess the distributions
of both parameters in the training dataset and those generated by using t-SNE  \cite{van2008visualizing} which produces a mapping of the parameter into a two-dimensional plane, as illustrated in Figure \ref{fig:9}. It can be observed that the proposed framework is capable of generating generalized parameters conditioned on the target graph data.

\begin{table}
 \caption{The average improvements of our proposed method against the baseline for random graphs in the format of
’original/generation’} \label{table_random}
  \centering
  \begin{tabular}{cccccc}
    \toprule
   Edge probability \textbackslash Node size    & 4 nodes & 5 nodes & 6 nodes & 7 nodes & 8 nodes   \\
    \midrule
    0.30 & 0.827/\textbf{0.956}  & 0.825/\textbf{0.916} & 0.855/\textbf{0.891} & 0.855/\textbf{0.885} & 0.854/\textbf{0.870}      \\
    0.35     & 0.837/\textbf{0.936}  & 0.824/\textbf{0.904} & 0.826/\textbf{0.899} & 0.811/\textbf{0.886} & 0.837/\textbf{0.874}      \\
    0.40    & 0.837/\textbf{0.926}  & 0.845/\textbf{0.911} & 0.807/\textbf{0.894} & 0.826/\textbf{0.902} & 0.838/\textbf{0.867}      \\
    0.45   & 0.875/\textbf{0.946}  & 0.859/\textbf{0.909} & 0.805/\textbf{0.903} & 0.893/\textbf{0.945} & 0.847/\textbf{0.900}      \\
    0.50     & 0.793/\textbf{0.937}  & 0.843/\textbf{0.918} & 0.815/\textbf{0.885} & 0.869/\textbf{0.937} & 0.860/\textbf{0.876}     \\
	0.55   & 0.844/\textbf{0.936}  & 0.824/\textbf{0.884} & 0.854/\textbf{0.909} & 0.861/\textbf{0.906} & 0.792/\textbf{0.880} \\
	0.60    & 0.863/\textbf{0.947}  & 0.820/\textbf{0.906} & 0.850/\textbf{0.896} & 0.784/\textbf{0.908} & 0.805/\textbf{0.902} \\
	0.65   & 0.867/\textbf{0.966}  & 0.852/\textbf{0.891} & 0.839/\textbf{0.907} & 0.799/\textbf{0.936} & 0.799/\textbf{0.904} \\
	0.70  & 0.849/\textbf{0.946}  & 0.788/\textbf{0.901} & 0.825/\textbf{0.887} & 0.847/\textbf{0.890} & 0.787/\textbf{0.882} \\
	0.75 & 0.813/\textbf{0.955}  & 0.830/\textbf{0.916} & 0.850/\textbf{0.887} & 0.845/\textbf{0.879} & 0.788/\textbf{0.891} \\
    \bottomrule
  \end{tabular}
  \label{tab:table}
\end{table}

\begin{table}
 \caption{The average improvements of our proposed method against the baseline for regular graphs in the format of 'original/generation'} \label{sec:table_r}
  \centering
  \begin{tabular}{cccccc}
    \toprule
   Node degree \textbackslash Node size    & 4 nodes & 5 nodes & 6 nodes & 7 nodes & 8 nodes   \\
    \midrule
    3 & 0.883/\textbf{0.900}  & - & 0.771/\textbf{0.853} & - & 0.763/\textbf{0.847}      \\
    4     & -  & 0.870/\textbf{0.900} & 0.844/\textbf{0.896} & 0.789/\textbf{0.870} & 0.763/\textbf{0.840}      \\
    5    & -  & - & 0.836/\textbf{0.900} & - & 0.737/\textbf{0.832}      \\
    6   & -  & - & - & 0.796/\textbf{0.858} & 0.766/\textbf{0.877}      \\
    7     & -  & - & - & - & 0.764/\textbf{0.835}     \\
    \bottomrule
  \end{tabular}
  \label{tab:table}
\end{table}

\begin{table}
 \caption{The average improvements of our proposed method against the baseline for Watts-Strogatz graphs in the format of 'original/generation'} \label{sec:table}
  \centering
  \begin{tabular}{cccccc}
    \toprule
   Edge probability \textbackslash Node size    & 4 nodes & 5 nodes & 6 nodes & 7 nodes & 8 nodes   \\
    \midrule
    0.30 & 0.837/\textbf{0.897}  & 0.835/\textbf{0.886} & 0.765/\textbf{0.879} & 0.821/\textbf{0.856} & 0.747/\textbf{0.829}      \\
    0.35     & 0.864/\textbf{0.898}  & 0.804/\textbf{0.879} & 0.777/\textbf{0.853} & 0.770/\textbf{0.836} & 0.747/\textbf{0.829}      \\
    0.40    & 0.810/\textbf{0.897}  & 0.835/\textbf{0.886} & 0.79/\textbf{0.852} & 0.808/\textbf{0.846} & 0.766/\textbf{0.798}      \\
    0.45   & 0.810/\textbf{0.897}  & 0.820/\textbf{0.882} & 0.801/\textbf{0.853} & 0.806/\textbf{0.851} & 0.787/\textbf{0.831}      \\
    0.50     & 0.846/\textbf{0.897}  & 0.776/\textbf{0.876} & 0.780/\textbf{0.848} & 0.810/\textbf{0.844} & 0.813/\textbf{0.859}     \\
	0.55   & 0.780/\textbf{0.896}  & 0.776/\textbf{0.876} & 0.798/\textbf{0.854} & 0.801/\textbf{0.847} & 0.791/\textbf{0.843} \\
	0.60    & 0.780/\textbf{0.896}  & 0.823/\textbf{0.882} & 0.782/\textbf{0.862} & 0.746/\textbf{0.828} & 0.763/\textbf{0.812} \\
	0.65   & 0.832/\textbf{0.896}  & 0.820/\textbf{0.872} & 0.808/\textbf{0.852} & 0.756/\textbf{0.824} & 0.772/\textbf{0.822} \\
	0.70  & 0.851/\textbf{0.897}  & 0.809/\textbf{0.874} & 0.821/\textbf{0.856} & 0.747/\textbf{0.829} & 0.785/\textbf{0.840} \\
    \bottomrule
  \end{tabular}
  \label{tab:Watts}
\end{table}

\begin{table}
 \caption{The average improvements of our proposed method against the baseline for graphs outside the training dataset regime in the format of 'original/generation'} \label{sec:table}
  \centering
  \begin{tabular}{cccc}
    \toprule
    Node size    & random graph & regular graph & Watts-Strogatz graph  \\
    \midrule
    9 & 0.696/\textbf{0.890}  & 0.780/\textbf{0.898} & 0.706/\textbf{0.781}  \\
    10     & 0.668/\textbf{0.835}  & 0.616/\textbf{0.900} & 0.651/\textbf{0.817}       \\
    11    & 0.626/\textbf{0.870}  & 0.820/\textbf{0.852} & 0.632/\textbf{0.787}      \\
    12   & 0.744/\textbf{0.877}  & 0.749/\textbf{0.841} & 0.749/\textbf{0.782}      \\
    13     & 0.712/\textbf{0.863}  & 0.683/\textbf{0.861} & 0.748/\textbf{0.780}      \\
    14   & 0.616/\textbf{0.809}  & 0.663/\textbf{0.829} & 0.748/\textbf{0.836}  \\
    15    & 0.808/\textbf{0.889}  & 0.692/\textbf{0.810} & 0.647/\textbf{0.834}  \\
    16   & 0.779/\textbf{0.873}  & 0.624/\textbf{0.857} & 0.793/\textbf{0.769}  \\
    \bottomrule
  \end{tabular}
  \label{tab:large}
\end{table}

\begin{figure}[htbp]
    \centering
    \includegraphics[width= 0.55 \textwidth]{t-SNE.png} 
    \caption{t-SNE visualizations of original training samples and generated parameters.} \label{fig:9}
\end{figure}

\section{Conclusion}\label{sec2}

In conclusion, delving into the commonalities between the diffusion-based generation process and QAOA's parameter training, we present a significant advancement in combining the conditional DDPM with QAOA,  specifically targeting the Max-Cut problem. To show the potential of the conditional DDPM in enhancing the initialization process of QAOA parameters, we conduct extensive experiments. The results demonstrate that the integration of the conditional DDPM with QAOA can improve the approximation ratio of QAOA compared to the baseline, Moreover, we also show that the proposed scheme can be extrapolated to larger Max-Cut problem instances, which are beyond the training data regime, and consequently reduce quantum computational resource overhead. This research establishes the foundation for future explorations in quantum computing, highlighting the diffusion model as a crucial instrument in this rapidly evolving
field.  

\bmhead{Acknowledgements}
This work is supported by the Jiangsu Funding Program for Excellent Postdoctoral Talent No.2022ZB139 and the Innovation Program for Quantum Science and Technology No. 2021ZD0302901.

\bibliography{sn-bibliography}

\end{document}